\documentclass[aps,floatfix,superscriptaddress,showpacs,twocolumn]{revtex4}
\usepackage{amsfonts}

\usepackage{graphicx}
\usepackage{array}
\usepackage{amsthm}
\usepackage{amsmath}
\usepackage{amssymb}
\usepackage{txfonts}

\let\mathbb=\varmathbb
\DeclareSymbolFont{letters}{OML}{ztmcm}{m}{it}

\input{tcilatex}

\begin{document}

\title{Dephasing-assisted parameter estimation in the presence of dynamical
decoupling}
\author{Qing-Shou Tan}
\affiliation{Zhejiang Institute of Modern Physics, Department of Physics, Zhejiang
University, Hangzhou 310027, China}
\author{Yixiao Huang}
\affiliation{Zhejiang Institute of Modern Physics, Department of Physics, Zhejiang
University, Hangzhou 310027, China}
\author{ Le-Man Kuang}
\affiliation{Key Laboratory of Low-Dimensional Quantum Structures and Quantum Control of
Ministry of Education, and Department of Physics, Hunan Normal University,
Changsha 410081, China}
\author{Xiaoguang Wang}
\affiliation{Zhejiang Institute of Modern Physics, Department of Physics, Zhejiang
University, Hangzhou 310027, China}

\begin{abstract}
We study the dephasing-assisted precision of parameter estimation (PPE)
enhancement in atom interferometer under dynamical decoupling (DD) pulses.
Through calculating spin squeezing (SS) and quantum Fisher information
(QFI), we find that dephasing noise can improve PPE by inducing SS, and the
DD pulses can maximize the improvement. It is indicated that in the presence
of DD pulses, the dephasing-induced SS can reach the limit of
\textquotedblleft one-axis twisting\textquotedblright\ model, $\xi^2\simeq
N^{-2/3}$ with $\xi^2$ being the SS parameter and $N$ the number of atoms.
In particular, we find that the DD pulses can amplify the dephasing-induced
QFI by a factor of $\simeq N/2$ compared with the noise-free case, which
means that under the control of DD pulses, the dephasing noise can enhance
the PPE to the scale of $\sqrt{2}/N$, the same order of magnitude of
Heisenberg limit ($1/N$).
\end{abstract}

\maketitle
\affiliation{Zhejiang Institute of Modern Physics, Department of Physics, Zhejiang
University, Hangzhou 310027, China}
\affiliation{Zhejiang Institute of Modern Physics, Department of Physics, Zhejiang
University, Hangzhou 310027, China}
\affiliation{Department of Physics, Hangzhou Normal University, Hangzhou 310036, China}
\affiliation{Zhejiang Institute of Modern Physics, Department of Physics, Zhejiang
University, Hangzhou 310027, China}
\email{xgwang@zimp.zju.edu.cn}

%03.65.Ta Foundations of quantum mechanics; measurement theory
%06.20.Dk Measurement and error theory
%06.20.-f Metrology
%03.65.Yz: Decoherence; open systems; quantum statistical methods

\address{Zhejiang Institute of Modern Physics, Department of Physics,
Zhejiang University, Hangzhou 310027, China}

%%%%%%%%%%%%%%%%%%%%%%%%%%%%%%%%%%%%%%

\section{Introduction}

Atom interferometry has attracted much attention because of its potential
applications in quantum metrology \cite{gus,fix,cro,gros,bar,rie,gro,mj}.
Atomic Bose-Einstein condensates (BECs) due to their unique coherence
properties and the possibility to yield controlled nonlinearity are viewed
as the ideal sources for an atom interferometer \cite{bou,kas}. The
nonlinearity of BECs caused by interatomic interactions can create squeezed
states \cite{ore,wine,ueda,jgr,pezz,YCL}, which can improve the precision of
parameter estimation (PPE).

The ability of BECs to create highly squeezed states and serve as nonlinear
interferometers, whose precision exceeds the standard quantum limit (SQL)
achieved with coherent spin states (CSS), has been demonstrated in two
recent experiments \cite{gros, rie}. However, the atom-atom nonlinear
interaction strength caused by s-wave scattering 
%,which creates spin squeezing (SS),% %
is usually very small when the modes of the BEC have a spatial overlap \cite%
{mj,gros,rie}. Such nonlinearity enhancement currently resorts to the use of
Feshbach resonances \cite{gros} or spatially separating the components of
BEC \cite{rie}, but the price of these methods is significantly increased
atom losses, limiting the achievable squeezing. In Ref. \cite{bar}, the
authors proposed an approach to drastically enhance the nonlinear dynamics
of the BEC based on collisions of the BEC with a thermal reservoir and
attained a strongly squeezing. This enhanced squeezing stems from the
decoherence noise which implies that the reservoir noise can also be
regarded as a resource to improve the parameter estimation sensitivity.
However, it is well known that the decoherence typically play a
coherence-destructive role which is one of the main obstacles to produce
certain spin-squeezed states (SSS). Much research showed that the
decoherence may prevent the production of certain SSS and limit the
precision of quantum metrology \cite{sim, and, sto, li, sin, wata, wang}.
Thus, it is important to suppress the coherence-destructive role but at the
same time maintain the decoherence-induced nonlinearity interaction if one
wants to obtain a strong squeezing and improve PPE.

Dynamical decoupling (DD) technique \cite{vio,kho, gord, uhr, yang, du, DJF,
GJB,pan,tan}, which has been widely employed in the area of quantum
information, provides an active way to fight against decoherence. Recently,
this technique has also been introduced into the field of magnetometers to
improve the sensitivity of oscillating magnetic fields based on
nitrogen-vacancy centers \cite{tay, lan, G.G,hall}. Combining the DD
technique with quantum metrology can preserve PPE in noisy system by
suppressing the decoherence effect \cite{tan}. Thus a natural question
rises, is it possible to realize decoherence-enhanced PPE under the DD
pulses?

In this paper, we give a positive answer to the above question by
investigating the influence of the DD pulses on the dephasing-induced SS and
dephasing-amplified quantum Fisher information (QFI) \cite%
{pezz,brau,hels,giova,sunz,ferr,maj,huang}, which are two important
quantities relevant in parameter estimation\cite{pezz,ferr}. We compare the
effects of two different DD pulse sequence: periodic DD (PDD) sequence \cite%
{vio, kho} and Uhrig DD (UDD) sequence \cite{uhr,yang}. Our finding shows
that both these sequences can effectively suppress the coherence-destructive
role and maintain the decoherence-induced nonlinearity interaction. It is
also found that the UDD sequence can work more efficiently, which can
enhance the decoherence-induced SS to the limit of $\xi ^{2}\simeq N^{-2/3}$ 
\cite{ueda, jgr} more easily, where $\xi ^{2}$ is the SS parameter and $N$
is the number of atoms. In particular, we find that it is possible for the
UDD sequence to amplify the QFI by a factor of $\simeq N/2$ compared with
the initial CSS in the case of pure dephasing. It means that the
dephasing-assisted sensitivity of the estimated parameter can be enhanced
from the SQL $1/\sqrt{N}$ to $\sqrt{2}/N$, which approaches nearly
Heisenberg-limited precision $(1/N)$ \cite{gio,hue}.

This paper is organized as follows. In Sec. II, we introduce our physical
model and Hamiltonian in the presence of control pulses. We then investigate
the dynamical evolution of the BEC system in the dephasing environment with
two different DD-pulse sequences. It is indicated that the DD-pulse
sequences can effectively suppress the coherence-destructive role while
maintaining the decoherence-induced nonlinearity interaction. In Sec. III,
we study the effects of DD pulses on enhancing the dephasing-induced spin
squeezing. We show that the magnitude of SS in the limit of $\xi ^{2}\simeq
N^{-2/3}$ can be induced by the dephasing noise in the presence of DD
pulses. Section IV discusses the dephasing-assisted QFI amplification under
the DD pulses. It is found that the DD pulses can greatly amplify the QFI
and enhance the PPE to the scale of Heisenberg limit. Finally, we conclude
this work in Sec. V.

\section{Dephasing in two-component Bose-Einstein condensate system with DD
pulse sequences}

In this section, we consider a two-component BEC system confined in a
harmonic potential, which suffers from dephasing noise.

\subsection{Model and Hamiltonian}

The total Hamiltonian is supposed to be 
\begin{equation}
H=\lambda \epsilon (s)J_{z}+\chi J_{z}^{2}+\sum_{k}\omega _{k}b_{k}^{\dag
}b_{k}+\epsilon (s)J_{z}\sum_{k}g_{k}\left( b_{k}^{\dag }+b_{k}\right) ,
\end{equation}%
which is the \textquotedblleft one-axis twisting\textquotedblright\ (OAT)
model in the presence of environment noise as well as DD control. Angular
momentum operators $J_{+}=(J_{-})^{\dag }=b^{\dag }a$ and $J_{z}=(b^{\dag
}b-a^{\dag }a)/2$ satisfy SU(2) algebra, with $a$ and $b$ being the
annihilation operators for two internal hyperfine states $\left\vert
a\right\rangle $ and $\left\vert b\right\rangle $ of the condensed atoms.
The nonlinear interaction strengthen $\chi $ can be controlled by using a
Feshbach resonance\cite{gros}. $\epsilon (s)$ is the time-dependent
modulation filed induced by $n$ DD $\pi $ pulses \cite{gord}, which reverse
the sign of $J_{z}$ at\ time $t_{j}$, and is given by 
\begin{equation}
\epsilon (s)=\sum_{j=0}^{n}(-1)^{j}\theta (s-t_{j})\theta (t_{j+1}-s),
\end{equation}%
with $s\in \lbrack 0,t].$ Here the total time interval $0\rightarrow t$ is
split into $n+1$ small intervals $t_{j}$ which satisfy $t_{0}=0$ and $%
t_{n+1}=t.$ In the above equation the step function $\theta (x)$ is equal to
1 if $x>0$ and 0 if $x<0$.

%In Eq. (1) the last term represents the interaction between the system and
%the reservoir with a coupling constant $g_{k}$, and the third term is the
%Hamiltonian of the reservoir.

In the interaction picture with respect to the reservoir operator $%
\sum_{k}\omega _{k}b_{k}^{\dag }b_{k}$, the time evolution operator can be
obtained by using Magnus expansion \cite{bar, bla}

\begin{equation}
U(t)=\mathrm{T}_{+}\exp \left[ -i\int_{0}^{t}H_{I}(t^{\prime })dt^{\prime }%
\right] =\exp \left[ iJ_{z}^{2}\Omega (t)\right] V(t),  \label{u1}
\end{equation}%
where the noise-induced nonlinear interaction strength can be recasted as 
\begin{equation}
\Omega (t)=\sum_{k}g_{k}^{2}\int_{0}^{t}ds\int_{0}^{s}ds^{\prime }\epsilon
(s)\epsilon (s^{\prime })\sin \omega _{k}(s-s^{\prime }).
\end{equation}%
In Eq. (3), the unitary operator $V(t)$ is defined by 
\begin{equation}
V(t)=e^{-i\int_{0}^{t}[\lambda \epsilon (s)J_{z}+\chi
J_{z}^{2}]ds}e^{J_{z}\sum_{k}(\alpha _{k}b_{k}^{\dag }-\alpha _{k}^{\ast
}b_{k})},  \label{v}
\end{equation}%
with the amplitudes $\alpha _{k}=-ig_{k}\int_{0}^{t}e^{i\omega
_{k}s}\epsilon (s)ds$. Furthermore, according to the experiment \cite{gros},
the nonlinear interaction strength $\chi $ is very small, $\chi \simeq 0$,
if without the Feshbach resonance. Thus we further have 
\begin{equation}
U(t)=e^{-i\phi }\exp [i\Omega (t)J_{z}^{2}]\exp \left[ J_{z}\sum_{k}(\alpha
_{k}b_{k}^{\dag }-\alpha _{k}^{\ast }b_{k})\right] ,
\end{equation}%
with $\phi =J_{z}\int_{0}^{t}\lambda \epsilon (s)ds$ \ which can be removed
when appropriate DD pulses sequences are applied, such as the UDD pulses
sequence and the odd number of PDD pulses sequence.

\subsection{System dynamical evolution under DD-pulse sequences}

In what follows, we investigate the dynamical evolution of the BEC system in
the dephasing environment with DD-pulse sequences. Let us assume that the
initial state of the total system is given by

\begin{equation}
\rho (0)=\left\vert \Psi (0)\right\rangle \left\langle \Psi (0)\right\vert
\otimes \rho _{B},  \label{inis}
\end{equation}%
where 
\begin{equation*}
\left\vert \Psi (0)\right\rangle =\sum_{m}c_{m}(0)\left\vert j,m\right\rangle
\end{equation*}%
is the CSS, with the probability amplitudes $c_{m}=2^{-j}\left(
C_{2j}^{j+m}\right) ^{1/2}$ and $\ $total spin $j=N/2$ for a system
consisting of $N$ atoms. Such a state is the optimal initial state to obtain
the strongest squeezing\cite{ueda,jgr}. In Eq. (\ref{inis}), $\rho _{B}$ is
the thermal equilibrium state of reservoir, defined by 
\begin{equation*}
\rho _{B}=\Pi _{k}[1-\exp (-\beta \omega _{k})]\exp (-\beta \omega
_{k}b_{k}^{\dag }b_{k}),
\end{equation*}%
with $\beta $ the inverse temperature ($\beta =1/T$).

Based on Eq. (5), the matrix elements of the system's density matrix can be
determined from the relation 
\begin{eqnarray}
\rho _{jm,jn}(t) &=&\mathrm{Tr}_{B}\left[ \left\langle j,m\right\vert
U(t)\rho (0)U^{-1}(t)\left\vert j,n\right\rangle \right]  \notag  \label{rou}
\\
&=&e^{-i(\phi _{m}-\phi _{n})}e^{i(m^{2}-n^{2})\Omega (t)}\exp \left[
-(m-n)^{2}R(t)\right] \rho _{jm,jn}(0).  \notag \\
&&
\end{eqnarray}%
In the above equation the decoherence function (see Appendix A for details) 
\begin{equation}
R(t)=\int_{0}^{\infty }d\omega \mathrm{F}(\omega ,t)G\text{ }(\omega )
\label{decf}
\end{equation}%
is the overlap integral of the temperature-dependent interacting spectrum%
\begin{equation*}
G(\omega )=J(\omega )[2n(\omega )+1]=J(\omega )\coth (\beta \omega /2),
\end{equation*}%
where $n(\omega )=[\exp (\beta \omega )-1]^{-1}$ is the bosonic distribution
function of the heat reservoir, $J(\omega )$ is the spectral density. The
filter function of an $n$-pulse sequence 
\begin{eqnarray*}
\mathrm{F}_{n}(\omega ,t) &\equiv &\frac{\left\vert \epsilon (\omega
)\right\vert ^{2}}{2}=\frac{1}{2}\left\vert \int_{0}^{t}e^{i\omega
s}\epsilon (s)ds\right\vert ^{2} \\
&=&\frac{1}{2\omega ^{2}}\left\vert 1+(-1)^{n+1}e^{i\omega
t}+2\sum_{j=1}^{n}(-1)^{j}e^{i\omega t_{j}}\right\vert ^{\substack{ 2}}.
\end{eqnarray*}%
Substituting the spectral density $J(\omega )$ into Eq. (4), the
noise-induced nonlinear term in Eq. (8) can be recasted as (see Appendix B ) 
\begin{equation}
\Omega (t)=\int_{0}^{\infty }d\omega J(\omega )f_{n}(\omega ,t),
\label{noise}
\end{equation}%
with 
\begin{eqnarray}
f_{n}(\omega ,t) &=&\mathrm{\vartheta }(\omega ,t)+\mathrm{\mu }(\omega
,t)+t/\omega ,  \notag  \label{filter} \\
\mathrm{\vartheta }(\omega ,t) &=&\frac{1}{\omega ^{2}}\left[ (-1)^{n+1}\sin
(\omega t)-2\sum_{m=1}^{n}(-1)^{m}\sin (\omega t_{m})\right] ,  \notag \\
\mathrm{\mu }(\omega ,t) &=&\frac{2}{\omega ^{2}}\left\{
\sum_{m=1}^{n}\sum_{j=1}^{m}(-1)^{m+j}\left( \sin [\omega
(t_{m}-t_{j})]\right. \right.  \notag \\
&&-\left. \left. \sin [\omega (t_{m+1}-t_{j})]\right) \right\} .
\end{eqnarray}%
Note that the result attained in the above equation is more complex than
that in Ref. \cite{uhr} for single-qubit DD.

For an ohmic bath in the continuum limit, 
\begin{equation*}
J(\omega )=\alpha \omega e^{-\omega /\omega _{c}}
\end{equation*}%
with $\alpha $ the coupling strength between the system and the reservoir
and $\omega _{c}$ the cutoff frequency, we find, in the absence of control $%
(n=0)$ \cite{lmk,hp} 
\begin{eqnarray}
R(t) &=&\alpha \int_{0}^{\infty }d\omega \omega e^{-\omega /\omega
_{c}}\coth (\beta \omega /2)\frac{1-\cos (\omega t)}{\omega ^{2}}, \\
\Omega (t) &=&\alpha \int_{0}^{\infty }d\omega \omega e^{-\omega /\omega
_{c}}\frac{\omega t-\sin (\omega t)}{\omega ^{2}}  \notag \\
&=&\alpha \lbrack \omega _{c}t-\arctan (\omega _{c}t)],
\end{eqnarray}%
where the decoherence function $R(t)$ can be further reduced as 
\begin{equation*}
R(t)=\alpha \left\{ \frac{1}{2}\ln (1+\omega _{c}^{2}t^{2})+\ln \left[ \frac{%
\beta }{\pi t}\sinh ({\pi t}/{\beta })\right] \right\}
\end{equation*}%
in the low temperature case ($\beta \omega _{c}\ll 1$).

If we consider the PDD sequence, in which sequence the $\pi $-pulse is
applied at equidistant intervals 
\begin{equation*}
t_{j}^{\mathrm{PDD}}=jt/(n+1),
\end{equation*}%
then the modulation spectrums $\mathrm{F}_{n}(\omega ,t)$ and $f_{n}(\omega
,t)$ in Eqs. (\ref{decf}) and (\ref{noise}) can be given by \cite{uhr} 
\begin{eqnarray}
\mathrm{F}_{n}^{\mathrm{PDD}}(\omega ,t) &=&\tan ^{2}[\omega
t/(2n+2)][1+(-1)^{n}\cos (\omega t)]/\omega ^{2},  \notag \\
f_{n}^{\mathrm{PDD}}(\omega ,t) &=&\frac{2(-1)^{n+1}\sin (\omega t)+\omega t%
}{\omega ^{2}}  \notag \\
&&+2\tan \left( \frac{\omega t}{2n+2}\right) \frac{(-1)^{n}\cos (\omega t)-n%
}{\omega ^{2}}  \notag \\
&&+\tan ^{2}\left( \frac{\omega t}{2n+2}\right) \frac{(-1)^{n}\sin (\omega t)%
}{\omega ^{2}}.
\end{eqnarray}%
Whereas for the UDD sequence \cite{uhr} 
\begin{equation}
t_{j}^{\mathrm{UDD}}=t\sin ^{2}\left[ {j\pi }/{(2n+2)}\right] ,  \notag
\end{equation}%
the filter function is

\begin{equation}
\mathrm{F}_{n}^{\mathrm{UDD}}(\omega ,t)\approx 8(n+1)^{2}J_{n+1}^{2}(\omega
t/2)/\omega ^{2},
\end{equation}%
where $J_{n}(x)$ is the Bessel function. The function $f_{n}^{\mathrm{UDD}%
}(\omega ,t)$ can be got by inserting $t_{j}^{\mathrm{UDD}}$ into Eq. (\ref%
{filter}). The form of the expression is very complex, and we do not give it
here.

\begin{figure}[tph]
\includegraphics[scale=0.6] {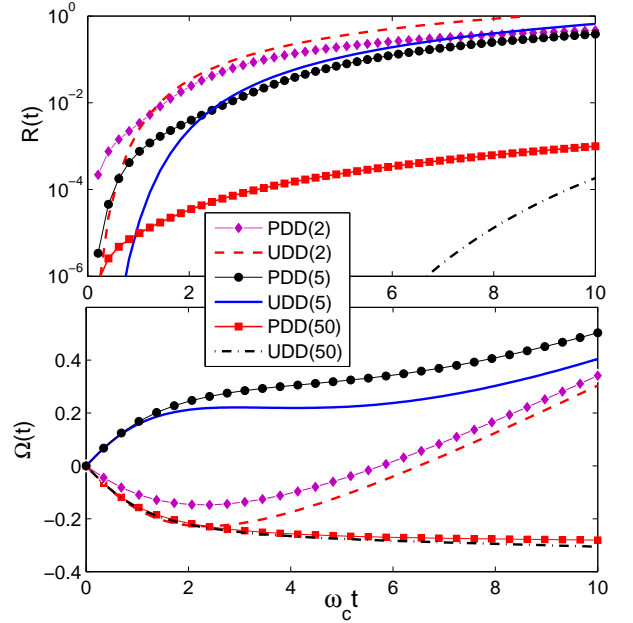} 
\caption{(Color online) Comparison of dynamical behaviors of functions $R(t)$
and $\Omega (t)$ under two different DD-pulse sequences. Relevant parameters
are chosen as coupling strength $\protect\alpha =0.1$ and temperature $T=%
\protect\omega _{c}$. }
\end{figure}
Although both the functions of $R(t)$ and $\Omega (t)$ stem from environment
noise, they play different roles. In other words, $\Omega (t)$ can induce
the quantum correlation in the system, while $R(t)$ destroys it. These
results imply that if one wants to make full use of the advantage of
environment noise to generate the desired quantum correlation, the
coherence-destructive process must be suppressed. Fortunately, according to
Eqs. (9)-(15) as well as Fig. 1, it is found that the DD-pulse can
effectively average the decoherence function $R(t)$ nearly to zero, but do
not remove the noise-induced nonlinear term $\Omega (t)$.

In the discussion below, we will investigate how to attain the best
squeezing and QFI, which are two quantities relevant in interferometry, in
the presence of dephasing by using of DD-pulse sequences.

\section{ Dephasing-induced spin squeezing in the present of DD-pulses
sequences}

In this section, we shall evaluate the magnitude of the dephasing-induced SS
as well as study how to improve it by employing the DD schemes as considered
above. To quantify the degree of SS, we introduce the SS parameter given by
Kitagawa and Ueda \cite{ueda} 
\begin{equation}
\xi ^{2}=\frac{2(\Delta J_{{\vec{n}}_{\bot }})_{\min }^{2}}{j},
\end{equation}%
where the minimization in the equation is over all directions denoted by $%
\vec{n}_{\bot }$, perpendicular to the mean spin direction. 
\begin{figure}[tph]
\includegraphics[scale=0.6] {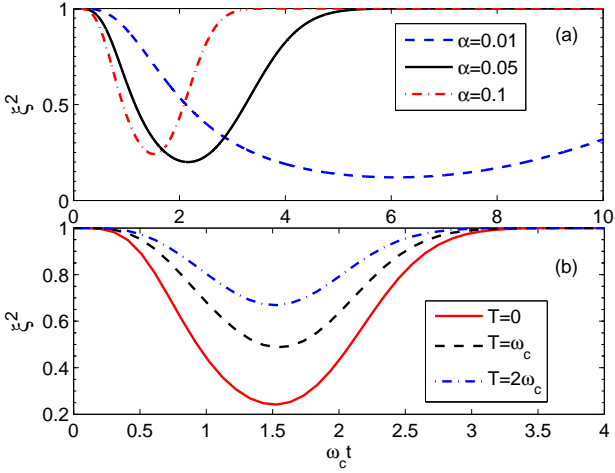} 
\caption{(Color online) Dephasing-induced spin squeezing in the absence of
control pulses ($n=0$). (a) Varying with different coupling strengthen $%
\protect\alpha $ with fixed $T=0$. (b) Varying with different values of
temperature with fixed $\protect\alpha =0.1$. The number of atoms is $N=200$%
. }
\end{figure}

With the use of Eq. (\ref{rou}), we can obtain the degree of
dephasing-induced SS for the initial state given in Eq. (\ref{inis}) \cite%
{ueda, jgr} in the case of pure dephasing noise 
\begin{equation}
\xi ^{2}=1+\frac{2j-1}{4}(A-\sqrt{A^{2}+B^{2}}),
\end{equation}%
in the optimally squeezed direction $\psi _{\text{opt}}=[\pi +\tan ^{-1}({B}/%
{A})]/2$, where%
\begin{eqnarray}
A &=&1-\cos ^{2j-2}[2\Omega (t)]e^{-4R(t)},  \notag \\
B &=&-4\sin [\Omega (t)]\cos ^{2j-2}[\Omega (t)]e^{-R(t)}.
\end{eqnarray}

Compared with Refs. \cite{ueda, jgr}, the controllable decoherence function $%
R(t)$ is introduced and the scaled time $\chi t$ is replaced by $\Omega (t)$
in the above equations. From the above equations, we can clearly find again
that dephasing noise plays two roles: on one hand, it can generate the SS by
inducing the nonlinear interaction $\Omega (t)$; on the other hand, it
degrade the degree of SS via the decoherence function $R(t)$. And from the
Eqs. (9), (14) and (15), it is found that if $t/(n+1)\rightarrow 0$, we have 
$R(t)\rightarrow 0$, then the coherence-destructive effect is suppressed.
Therefore, in the short-time limit ($\Omega (t)\ll 1$) and the large
particle number limit ($j\gg 1$), the best squeezing can be approximated as 
\begin{equation}
\xi _{\min }^{2}\simeq \frac{3}{4j}(\frac{2j}{3})^{1/3}\simeq N^{-2/3},
\label{min}
\end{equation}%
which is the well known result appeared in Ref. \cite{ueda, jgr} for an
ideal noise-free case.

In Fig. 2, we plot the dynamics of dephasing-induced SS in the absence of
control pulses. From Fig. 2 we can find that the decoherence not only
generate SS but also prevent the production of certain SS; and the stronger
coupling strength [Fig. 1(a)] or the higher the temperature [Fig. 1(b)] is,
the weaker the squeezing is attained. In all these cases the best squeezing
given in Eq. (\ref{min}) cannot be reached. 
\begin{figure}[tph]
\includegraphics[scale=0.6]{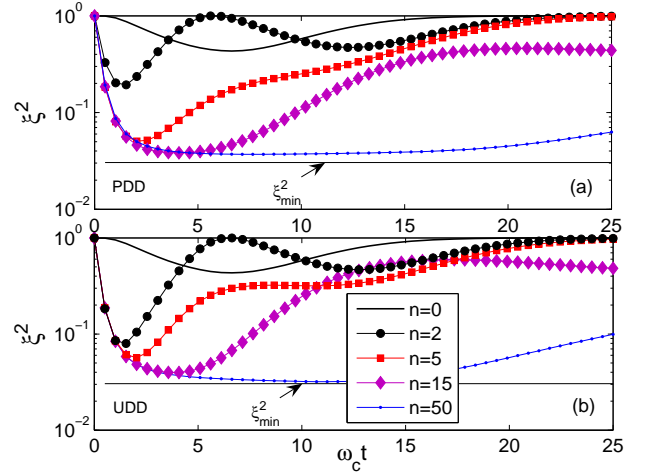}
\caption{(Color online) Spin squeezing $\protect\xi ^{2}$ with respect to
scaled time $\protect\omega _{c}t$ for (a) PDD sequence and (b) UDD sequence
with different numbers of control pulse $n$. Relevant parameters are chosen
as $\protect\alpha =0.01,T=\protect\omega _{c}$, and $N=200$. Here $\protect%
\xi _{\text{min}}^{2}$ is the approximations given in Eq. (\protect\ref{min}%
).}
\end{figure}

To clearly observe the effects of the DD pulses on SS, a comparison of the
consequence of the two different DD schemes on the dynamics of SS is given
in Fig. 3. It indicates that both these DD sequences can effectively improve
the magnitude of squeezing, and the UDD pulses work more efficiently than
the PDD pulses when they are used to enhance the decoherence-induced SS.
From Fig. 3(b), we can see that the strongest squeezing $\xi _{\min }^{2}$
for OAT given in Eq. (\ref{min}) can be obtained if the UDD sequences are
applied. This is an interesting phenomenon since this squeezing limit is
usually thought can be achieved only in the ideal noise-free OAT case, but
Fig. 3 shows that the noise also can induce it in the presence of DD-pulse
sequences.

It is well known that SS is an useful resource to improve PPE. Dephasing can
induce the squeezing, which means that the dephasing noise also can be
regarded as a resource to enhance the PPE sometimes rather than reduce it.

%%%%%%%%%%%%%%%%%%%%%%%%%%%%%%%%

\section{ Dephasing-Assisted QFI amplification in the presence of DD pulses}

To understand well the behaviors of dephasing-assisted enhancement of
sensitivity, we evaluate the QFI $\mathcal{F}$, which gives a
theoretical-achievable limit on the precision of an unknown parameter $%
\theta $ via Cram\'{e}r-Rao bound

\begin{equation*}
\Delta \theta \geq \Delta \theta _{\mathrm{QCR}}=\frac{1}{\sqrt{N_{m}%
\mathcal{F}}},
\end{equation*}%
with $N_{m}$ the number of measurements. Below, we set $N_{m}=1\ $for
simplicity.

According to Refs. \cite{mj,ferr,sunz,maj,huang}, the QFI $\mathcal{F}$ with
respect to $\theta ,$ acquired by an SU(2) rotation, can be explicitly
derived as

\begin{equation}
\mathcal{F}[\rho (\theta ,t),J_{\vec{n}}]=\mathrm{Tr}[\rho (\theta
,t)L_{\theta }^{2}]=\vec{n}\mathbf{C}\vec{n}^{T},  \label{fisher}
\end{equation}%
where 
\begin{equation*}
\rho (\theta ,t)=\exp (-i\theta J_{\vec{n}})\rho (t)\exp (i\theta J_{\vec{n}%
})
\end{equation*}%
and the matrix element for the symmetric matrix $\mathbf{C}$ is 
\begin{equation}
\mathbf{C}_{kl}=\sum_{i\neq j}\frac{(p_{i}-p_{j})^{2}}{p_{i}+p_{j}}%
[\left\langle i\right\vert J_{k}\left\vert j\right\rangle \left\langle
j\right\vert J_{l}\left\vert i\right\rangle +\left\langle i\right\vert
J_{l}\left\vert j\right\rangle \left\langle j\right\vert J_{k}\left\vert
i\right\rangle ],
\end{equation}%
where $p_{i}$($\left\vert i\right\rangle $) are the eigenvalues
(eigenvectors) of $\rho (\theta ,t).$

In particular, if $\rho $ is a pure state, the above matrix can be
simplified as \cite{ferr,sunz,maj,huang} 
\begin{equation}
\mathbf{C}_{kl}=2\left\langle J_{k}J_{l}+J_{l}J_{k}\right\rangle
-4\left\langle J_{k}\right\rangle \left\langle J_{l}\right\rangle .
\label{cmat}
\end{equation}%
From Eq. (\ref{fisher}), one finds that to get the highest possible
estimation precision $\Delta \theta $, a proper direction $\vec{n}$ should
be chosen for a given state, which maximizes the value of the QFI. With the
help of the symmetric matrix, the maximal mean QFI can be obtained as 
\begin{equation}
\mathcal{F}_{\max }=\lambda _{\max },  \label{max}
\end{equation}%
where $\lambda _{\max }$ is the maximal eigenvalues of $\mathbf{C}$. And for
the initial CCS we have $\mathcal{F}_{\mathrm{CSS}}=N$, then $\Delta \theta
_{\min }=1/\sqrt{N}$, which reaches SQL.

\begin{figure}[tph]
\includegraphics[scale=0.62] {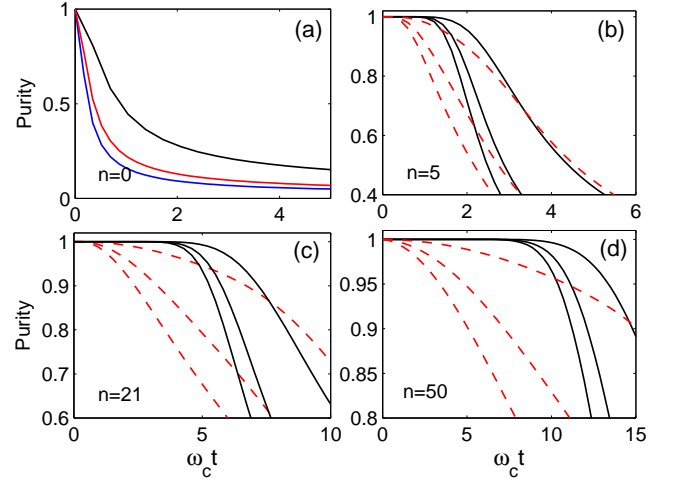} 
\caption{(Color online) Purity defined in Eq. (\protect\ref{purity}) vs time
for different number of DD pulse at $T=\protect\omega_c$ and $N=200$. Solid
lines are for the UDD sequence and dashed lines for the PDD sequence. From
the bottom to the top, the curves correspond to $\protect\alpha=0.1, 0.05$
and $0.01$.}
\end{figure}

As discussed in the previous sections, the DD pulses can decouple the state
of the system from environment by averaging the decoherence function $R(t)$
to zero. To further check its consequence, here we introduce the purity of a
quantum state, which is defined by 
\begin{equation}
P(\rho )\equiv \mathrm{Tr}(\rho ^{2}).  \label{purity}
\end{equation}%
The quantum state is pure iff its purity takes the maximum value 1, while it
is the maximally mixed state $\rho _{m}\equiv \mathbb{I}/D$ iff its purity
takes the minimum value $1/D,$ with $D$ the dimensional of quantum system 
\cite{tana}.

A comparison of the effects between UDD and PDD sequences for different
values of $\alpha $ on protecting the purity of the quantum state is shown
in Fig. 4.

Figure 4 indicates the advantage of the UDD sequence in preserving the the
purity of quantum state. In contrast to PDD pulses the UDD pulses can
preserve the maximum value of purity for a longer time and the behavior for
different couplings $\alpha $ do not change as obvious as the case of PDD
pulses. This feature of UDD-pulse means it has a long preservation time for
maximal purity even with large coupling. Thus, we can obtain a pure state $%
[P(\rho )=1]$ at certain times if the UDD pulses are applied. It is also
indicated that larger number of UDD pulses conduces longer preservation time
of pure state.

For a pure state [i.e., $R(t)=0$], based on Eqs. (\ref{cmat}) and (\ref{max}%
) the maximal QFI can be derived as (see Appendix C) 
\begin{equation}
\mathcal{F}_{\max }=N\eta (N,t),  \label{pure}
\end{equation}%
where 
\begin{eqnarray}
\eta (N,t) &=&\max \left\{ 1+\frac{N-1}{4}\left( A_{+}^{\prime }+\sqrt{%
A_{+}^{\prime 2}+B^{\prime 2}}\right) \right. ,  \notag  \label{rate} \\
&&\left. 1+\frac{N-1}{2}A_{-}^{\prime }-N\cos ^{2N-2}[\Omega (t)]\right\} ,
\end{eqnarray}%
with 
\begin{eqnarray}
A_{\pm }^{\prime } &=&1\pm \cos ^{N-2}[2\Omega (t)],  \notag \\
B^{\prime } &=&-4\sin [\Omega (t)]\cos ^{N-2}[\Omega (t)].
\end{eqnarray}%
\begin{figure}[tph]
\includegraphics[scale=0.62] {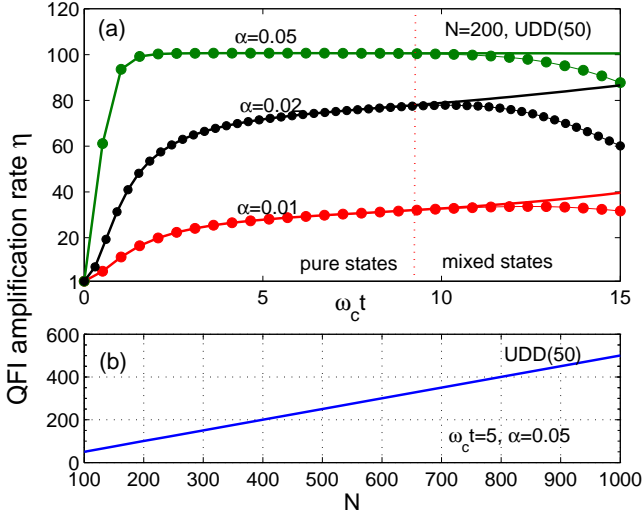} 
\caption{(Color online) (a) QFI amplification rate $\protect\eta $ vs the
scaled time $\protect\omega _{c}t$ for different values of coupling strength 
$\protect\alpha $ with the number of atoms $N=200$. The solid lines are
analytical results of pure state, while solid circles line present numerical
results of Eq. (\protect\ref{fisher}). In the pure states regime they fit
well. (b) QFI amplification rate as a function of atom number $N$ at fixed
time $\protect\omega _{c}t=5$ with $\protect\alpha =0.05$. The temperature
is set as $T=\protect\omega _{c}$ and the number of UDD pulse is $n=50$. }
\end{figure}

The above equations imply that the maximal QFI can be amplified $\eta $
times compared with CSS in the case of pure states.

In Fig. 5(a), we compare the QFI amplification rate $\eta$ versus scaled
time $\omega _{c}t$ between analytical results of pure states and numerical
results of actual states with fixed UDD pulse $(n=50)$. It can be seen from
Fig. 5(a) that the analytical results are good agreed with the numerical
results in the pure states regimes shown in Fig. 4. The amplification is by
a factor of 100 for the initial state (CSS) when $\alpha =0.05$ and $N=200$.
For $\alpha =0.01$ the amplification is still about a factor of 30. It is
indicated that large amplification rate can be easily reached for large $%
\alpha \ $in the presence of UDD pulses. Another interesting behavior is
that the large $\alpha $ (such as $\alpha =0.05$) can maintain the QFI
unchanged until the quantum state into the mixed states regimes. In the
mixed states regimes the amplification rate is reduced, which implies that a
larger number of UDD pulses is needed to extend the pure states preservation
time, if one wants to maintain the maximal steady amplification rate.

In Fig. 5(b), we plot the QFI amplification rate $\eta$ as a function of
atom number $N$ at fixed scaled time $\omega _{c}t=5$. Figure 5(b) shows
that the amplification is proportional to the atom number $N$, and the scale
factor is $\simeq 1/2$. It indicates that the amplification rate given in
Eq. (\ref{rate}) has the maximum value $\eta _{\max }(N)\simeq N/2,$ thus
the maximal QFI $\mathcal{F}_{\max }\simeq N^{2}/2$ in this case.

According to the quantum Cram\'{e}r-Rao theorem, we know that the larger QFI
is, the higher precision of estimation is obtained. Thus, the
dephasing-induced amplified QFI can greatly improve the the parameter
estimation precision; the best result is that it can enhance the phase
sensitivity from SQL $\Delta \theta =1/\sqrt{N}$ to $\Delta \theta =\sqrt{2}%
/N$, which is the same order of magnitude of Heisenberg limit $(1/N)$.

%%%%%%%%%%%%%%%%%%%%%%%%%%%%%%%%%%%%%%

\section{conclusion}

%%%%%%%%%%%%%%%%%%%%%%%%%%%%%%%%%%%%%%%

In summary, we have studied the dephasing-assisted PPE enhancement in a
two-component BEC system in the presence of DD pulses, through calculating
dephasing-induced SS and QFI. It has been found that the dephasing noise can
improve PPE by inducing SS. And the DD pulses can maximize the improvement.
We have compared the effects between PDD sequence and UDD sequence. Our
results showed that the UDD sequence can work more efficiently, which can
enhance the decoherence-induced SS to the limit of $\xi ^{2}\simeq N^{-2/3}$
more easily and it can amplify the QFI by a factor of $\simeq N/2$ for the
initial state of CSS. It implied that the sensitivity $\Delta \theta $ of
the estimated parameter $\theta $ can be enhanced from the SQL $1/\sqrt{N}$
to $\sqrt{2}/N$, which achieves nearly Heisenberg-limited precision $(1/N)$.

We would like to mention that, under our consideration, all the $\pi $
control pulses are assumed to execute quickly and perfect, during which the
coupling with environment is negligible. Besides, it should be pointed out
that, we also have noted that the Heisenberg-limited precision estimation
precision have been reached theoretically, via transforming the OAT into
\textquotedblleft two-axis twisting\textquotedblright\ \cite{YCL,GJB} if
some more complex control fields are employed. Finally, we expect that our
idea might have promising application in quantum metrology and be realized
within current experiments.

%\paragraph*{Acknowledgments}

\acknowledgments X. Wang acknowledges support from the NFRPC through Grant
No. 2012CB921602 and the NSFC through Grants No. 11025527 and No. 10935010.
L. M. Kuang acknowledges support from the 973 Program under Grant No.
2013CB921804, the NSF under Grant No. 11075050, the PCSIRTU under Grant No.
IRT0964, and the HPNSF under Grant No. 11JJ7001. 
%%%%%%%%%%%%%%%%%%%%%%%%%%%%%%%%%%%%%%%%%%%

\appendix{}

\section{derivation of the decoherence function $R(t)$}

Here, we recap the derivation of the decoherence function $R(t)$ for
completeness.

From Eq. (5), the reduced density matrix elements of the system can be read
as 
\begin{eqnarray}
\rho _{m,n}(t) &=&\mathrm{Tr}_{B}\left[ \left\langle m\right\vert U(t)\rho
(0)U^{-1}(t)\left\vert n\right\rangle \right]  \notag \\
&=&\mathrm{Tr}_{B}\left[ \left\langle m\right\vert e^{-i\phi
_{m}}e^{im^{2}\Omega (t)}\exp \left( m\sum_{k}(\alpha _{k}b_{k}^{\dag
}-\alpha _{k}^{\ast }b_{k})\right) \right.  \notag \\
&\times &\left. \rho (0)e^{i\phi _{n}}e^{-in^{2}\Omega (t)}\exp \left(
-n\sum_{k}(\alpha _{k}b_{k}^{\dag }-\alpha _{k}^{\ast }b_{k})\right)
\left\vert n\right\rangle \right]  \notag \\
&=&e^{-i(\phi _{m}-\phi _{n})}e^{i(m^{2}-n^{2})\Omega (t)}  \notag \\
&\times &\mathrm{Tr}_{B}\left\{ \exp \left[ (m-n)\sum_{k}(\alpha
_{k}b_{k}^{\dag }-\alpha _{k}^{\ast }b_{k})\right] \rho _{B}(0)\right\} \rho
_{m,n}(0)  \notag \\
&&
\end{eqnarray}%
with $\alpha _{k}=-ig_{k}\int_{0}^{t}e^{i\omega _{k}s}\epsilon (s)ds.$

To attain the explicit expression of the above equation, the main task
becomes to calculate the expectation value of displacement operator 
\begin{eqnarray}
\Pi _{k}\mathrm{Tr}_{B}\left[ D(z_{k})\rho _{B}\right] &=&\mathrm{Tr}%
_{B}\left\{ \exp \left[ (m-n)\sum_{k}(\alpha _{k}b_{k}^{\dag }-\alpha
_{k}^{\ast }b_{k})\right] \rho _{B}\right\}  \notag \\
&&
\end{eqnarray}%
with $z_{k}=(m-n)\alpha _{k}.$

Making use of the following formula \cite{hp} 
\begin{equation*}
\mathrm{Tr}_{B}\left[ D(z_{k})\rho _{B}\right] =\exp [-(\left\langle
n_{k}\right\rangle +1/2)\left\vert z_{k}\right\vert ^{2}],
\end{equation*}%
where $n_{k}=1/(e^{\beta \omega _{k}}-1)$.

Then we arrive at 
\begin{eqnarray}
\left\langle D(z_{k})\right\rangle &=&\mathrm{Tr}_{B}\left[ D(z_{k})\rho _{B}%
\right]  \notag \\
&=&\exp \left[ -(m-n)^{2}\left\vert -ig_{k}\int_{0}^{t}e^{i\omega
_{k}s}\epsilon (s)ds\right\vert ^{2}(\left\langle n_{k}\right\rangle +1/2)%
\right]  \notag \\
&=&\exp \left[ -(m-n)^{2}\left\vert g_{k}\right\vert ^{2}\left\vert
\int_{0}^{t}e^{i\omega _{k}s}\epsilon (s)ds\right\vert ^{2}(\left\langle
n_{k}\right\rangle +1/2)\right]  \notag \\
&=&\exp \left\{ -(m-n)^{2}\int_{0}^{\infty }d\omega J(\omega )[2n(\omega
)+1]\left\vert \epsilon (\omega )\right\vert ^{2}/2\right\}  \notag \\
&=&\exp \left[ -(m-n)^{2}R(t)\right] .
\end{eqnarray}

In the above equation the finite-time Fourier transform $\epsilon (\omega
,t) $ of $\epsilon (s)$ can be derived as

\begin{eqnarray}
\epsilon (\omega ,t) &=&\int_{0}^{t}e^{i\omega s}\epsilon (s)ds  \notag \\
&=&\sum_{j=0}^{n}\int_{0}^{t}e^{i\omega s}(-1)^{j}\theta (s-t_{j})\theta
(t_{j+1}-s)ds  \notag \\
&=&\sum_{j=0}^{n}(-1)^{j}\int_{t_{j}}^{t_{j+1}}e^{i\omega s}ds=\frac{1}{%
i\omega }\sum_{j=0}^{n}(-1)^{j}[e^{i\omega t_{j+1}}-e^{i\omega t_{j}}] 
\notag \\
&=&\frac{i}{\omega }\left[ 1+(-1)^{n+1}e^{i\omega
t}+2\sum_{j=1}^{n}(-1)^{j}e^{i\omega t_{j}}\right] .
\end{eqnarray}%
Therefore, the decoherence function in Eq. (8) is obtained.

\begin{widetext}
\section{derivation of $f(\omega,t)$}
In this Appendix, we present details of derivation of Eq. (10). From Eqs.
(4) and (10), we have
\begin{eqnarray}
f(\omega ,t) =\int_{0}^{t}ds\int_{0}^{s}ds^{\prime }\epsilon (s)\epsilon
(s^{\prime })\sin [\omega (s-s^{\prime })]
=\frac{1}{2i}[x(\omega ,t)-x^{\ast }(\omega ,t)]=\mathrm{Im}[x(\omega ,t)].
\end{eqnarray}
Thus, we obtain
\begin{eqnarray}
x(\omega ,t) &=&\int_{0}^{t}ds\epsilon (s)e^{i\omega
s}\int_{0}^{s}ds^{\prime }\epsilon (s^{\prime })e^{-i\omega s^{\prime
}}=\sum_{m=0}^{n}\int_{t_{m}}^{t_{m+1}}ds(-1)^{m}e^{i\omega
s}\int_{0}^{s}ds^{\prime }\epsilon (s^{\prime })e^{-i\omega s^{\prime }}
\notag \\
&=&\sum_{m=0}^{n}\int_{t_{m}}^{t_{m+1}}ds(-1)^{m}e^{i\omega s}\left[
\sum_{j=1}^{m}\int_{t_{j-1}}^{t_{j}}(-1)^{j-1}e^{-i\omega s^{\prime
}}ds^{\prime }+(-1)^{m}\int_{t_{m}}^{s}e^{-i\omega s^{\prime }}ds^{\prime }%
\right]  \notag\\
&=&-\frac{i}{\omega }\left\{
\sum_{m=0}^{n}\int_{t_{m}}^{t_{m+1}}ds(-1)^{m}e^{i\omega s}\left[
1+2\sum_{j=1}^{m}(-1)^{j}e^{-i\omega t_{j}}+(-1)^{m+1}e^{-i\omega s}\right]
\right\}   \notag \\
&=&-\frac{i}{\omega }\left\{
\sum_{m=0}^{n}\int_{t_{m}}^{t_{m+1}}ds(-1)^{m}e^{i\omega s}\left[
1+2\sum_{j=1}^{m}(-1)^{j}e^{-i\omega t_{j}}\right] +\sum_{m=0}^{n}%
\int_{t_{m}}^{t_{m+1}}ds(-1)^{2m+1}\right\}   \notag \\
&=&\frac{1}{\omega ^{2}}\left[ 1+(-1)^{n+1}e^{i\omega
t}+2\sum_{m=1}^{n}(-1)^{m}e^{-i\omega t_{m}}\right] -\frac{2i}{\omega }%
\sum_{m=1}^{n}\sum_{j=1}^{m}\int_{t_{m}}^{t_{m+1}}ds(-1)^{m+j}e^{i\omega
s}e^{-i\omega t_{j}}+\frac{it}{\omega }  \notag \\
&=&\Theta (\omega ,t)+\Xi (\omega ,t)+\frac{it}{\omega },
\end{eqnarray}%
where
\begin{eqnarray}
\Theta (\omega ,t)& =&\frac{1}{\omega ^{2}}\left[ 1+(-1)^{n+1}e^{i\omega
t}+2\sum_{m=1}^{n}(-1)^{m}e^{-i\omega t_{m}}\right] ,  \notag \\
\Xi (\omega ,t)&=&-\frac{2}{\omega ^{2}}\sum_{m=1}^{n}%
\sum_{j=1}^{m}(-1)^{m+j}e^{-i\omega t_{j}}(e^{i\omega t_{m+1}}-e^{i\omega
t_{m}}).
\end{eqnarray}%
Then we have
\begin{equation}
f(\omega ,t)=\mathrm{Im}[x(\omega ,t)]=\mathrm{\vartheta }(\omega ,t)+%
\mathrm{\mu }(\omega ,t)+t/\omega ,
\end{equation}%
where
\begin{equation}
\mathrm{\vartheta }(\omega ,t)=\mathrm{Im}[\Theta (\omega ,t)],\hspace{0.5cm}%
\mathrm{\mu }(\omega ,t)=\mathrm{Im}[\Xi (\omega ,t)],  \notag
\end{equation}%
which have been given in Eq. (10).
\end{widetext}

\section{QFI of pure states}

Based on Eq. (\ref{cmat}) of the main text, we have 
\begin{equation}
\mathbf{C}=4\left( 
\begin{array}{ccc}
(\Delta J_{x})^{2} & \mathrm{Cov}(J_{x},J_{y}) & \mathrm{Cov}(J_{x},J_{z})
\\ 
\mathrm{Cov}(J_{x},J_{y}) & (\Delta J_{y})^{2} & \mathrm{Cov}(J_{y},J_{z})
\\ 
\mathrm{Cov}(J_{x},J_{z}) & \mathrm{Cov}(J_{y},J_{z}) & (\Delta J_{z})^{2}%
\end{array}%
\right),
\end{equation}%
with $\mathrm{Cov}(J_{m},J_{n})=\frac{1}{2}\left\langle
J_{m}J_{n}+J_{n}J_{m}\right\rangle -\left\langle J_{m}\right\rangle
\left\langle J_{n}\right\rangle $.

When the UDD pulses are employed, we have $\phi =J_{z}\int_{0}^{t}\lambda
\epsilon (s)ds=0$ in Eq. (7). Following Eq. (7) and Ref. \cite{jgr}, the
expectation values relevant in $\mathrm{Cov}(J_{m},J_{n})$ can be attained
as 
\begin{eqnarray}
\langle J_{x}J_{y}+J_{y}J_{x}\rangle &=&\mathrm{Im}\langle J_{+}^{2}\rangle
=0,  \notag \\
\langle J_{x}J_{z}+J_{z}J_{x}\rangle &=&\mathrm{Re}\langle
J_{+}(2J_{z}+1)\rangle =0,  \notag \\
\langle J_{y}J_{z}+J_{z}J_{y}\rangle &=&\mathrm{Im}\langle
J_{+}(2J_{z}+1)\rangle ,
\end{eqnarray}%
with 
\begin{eqnarray*}
&&\langle J_{+}(2J_{z}+1)\rangle =i2j(j-1/2)\cos ^{2j-2}(2\Omega )\sin
(\Omega ), \\
&&\langle J_{+}^{2}\rangle =j(j-1/2)\cos ^{2j-2}(2\Omega ),\hspace{0.2cm}%
\langle J_{+}\rangle =j\cos ^{2j-1}(\Omega ),
\end{eqnarray*}%
and 
\begin{eqnarray}
\langle J_{x}\rangle &=&j\cos ^{2j-1}(\Omega ),\hspace{0.2cm}\langle
J_{y}\rangle =\langle J_{z}\rangle =0,\hspace{0.2cm}\langle J_{z}^{2}\rangle
=j/2,  \notag \\
\langle J_{x}^{2}\rangle &=&\frac{j}{4}(2j+1)+\frac{j}{4}(2j-1)\cos
^{2j-2}(2\Omega ),  \notag \\
\langle J_{y}^{2}\rangle &=&\frac{j}{4}(2j+1)-\frac{j}{4}(2j-1)\cos
^{2j-2}(2\Omega ).
\end{eqnarray}%
Hence, the symmetric matrix $\mathbf{C}$ can be rewritten as 
\begin{equation}
\mathbf{C}=4\left( 
\begin{array}{ccc}
(\Delta J_{x})^{2} & 0 & 0 \\ 
0 & \langle J_{y}^{2}\rangle & \mathrm{Cov}(J_{y},J_{z}) \\ 
0 & \mathrm{Cov}(J_{y},J_{z}) & \langle J_{z}^{2}\rangle%
\end{array}%
\right)
\end{equation}%
with the maximal eigenvalue 
\begin{equation}
\lambda _{\max }=4\max \{(\Delta J_{x})^{2},\lambda _{\pm }\},
\end{equation}%
where 
\begin{eqnarray*}
(\Delta J_{x})^{2} &=&\frac{N}{4}\left( \frac{N+1}{2}+\frac{N-1}{2}\cos
^{2j-2}(2\Omega )-N\cos ^{4j-2}(\Omega )\right) , \\
\lambda _{\pm } &=&\frac{\left\langle J_{y}^{2}+J_{z}^{2}\right\rangle \pm 
\sqrt{\left( \left\langle J_{y}^{2}+J_{z}^{2}\right\rangle \right) ^{2}+4%
\mathrm{Cov}(J_{y},J_{z})^{2}}}{2}.
\end{eqnarray*}

\end{document}